# Detection of microgauss coherent magnetic fields in a galaxy five billion years ago

S. A. Mao[1]*, C. Carilli[2,3], B. M. Gaensler[4], O. Wucknitz[1], C. Keeton[5], A. Basu[1], R. Beck[1], P. P. Kronberg[6], E. Zweibel[7,8]

[1]Max Planck Institute for Radio Astronomy, Auf dem Hügel 69, Bonn D-53121, Germany.

[2]National Radio Astronomy Observatory, P. O. Box O, Socorro, NM 87801, USA.

[3]Cavendish Astrophysics Group, Cambridge CB3 0HE, UK.

[4]Dunlap Institute for Astronomy & Astrophysics, University of Toronto, Toronto, ON, M5S 3H4, Canada.

[5]Department of Physics and Astronomy, Rutgers University, Piscataway, NJ 08854, USA.

[6]Department of Physics, University of Toronto, Toronto, ON, M5S 1A7, Canada.

[7]Department of Astronomy, The University of Wisconsin, Madison, WI 53706, USA.

[8]Department of Physics, The University of Wisconsin, Madison, WI 53706, USA.

*Correspondence to: mao@mpifr-bonn.mpg.de

**Magnetic fields play a pivotal role in the physics of interstellar medium in galaxies[1], but there are few observational constraints on how they evolve across cosmic time[2-7]. Spatially resolved synchrotron polarization maps at radio wavelengths reveal well-ordered large-scale magnetic fields in nearby galaxies[1,8,9] that are believed to grow from a seed field via a dynamo effect[10,11]. To directly test and characterize this theory requires magnetic field strength and geometry measurements in cosmologically distant galaxies, which are**

**challenging to obtain due to the limited sensitivity and angular resolution of current radio telescopes. Here, we report the cleanest measurements yet of magnetic fields in a galaxy beyond the local volume, free of the systematics traditional techniques would encounter. By exploiting the scenario where the polarized radio emission from a background source is gravitationally lensed by a foreground galaxy at $z = 0.439$ using broadband radio polarization data, we detected coherent μG magnetic fields in the lensing disk galaxy as seen 4.6 Gyrs ago, with similar strength and geometry to local volume galaxies. This is the highest redshift galaxy whose observed coherent magnetic field property is compatible with a mean-field dynamo origin.**

We measured the magnetic field in a $z = 0.439$ galaxy using Karl G. Jansky Very Large Array broadband (1-8 GHz) polarization observations of the strong lensing system CLASS B1152+199. The polarized background source at $z = 1.019$ was lensed by the late-type galaxy into two images (A and B) separated by 1.56" at impact parameters 6.5 and 2.6 kpc (refs [12,13]) as shown in Fig. 1. The lensing galaxy has been classified as a late-type star-forming disk galaxy with a substantial interstellar medium (ISM), based on strong O[II] emission, and large hydrogen, MgII, and dust column densities seen towards the lensed quasar images[12-15]. Characterization of the lensing galaxy enables one to establish links between its galaxy properties and magnetic fields. This is not achievable by standard Faraday rotation measure grid studies (inference of magnetic fields in distant galaxies using Faraday rotation—the integral of the line-of-sight magnetic field weighted by thermal electron density—of polarized radio sources behind them) based on samples of intervening galaxies of vastly different masses, types and viewing geometries. These studies have yielded only statistical detections of magnetic fields associated with cosmologically distant galaxies[3-7] without reliable field strength and geometry estimations.

Due to the achromatic and non-polarizing nature of lensing, the Faraday rotation and wavelength-dependent beam depolarization characterized by Faraday dispersion ($\sigma_{RM}$ describes the fluctuations in the magnetic field and electron density on scales smaller than the telescope beam in a foreground medium, which lead to cancellation of the polarization vectors of neighbouring sightlines within a beam, reducing the net measured polarization) of the two lensed images should remain unchanged unless there are propagation effects along ray bundles through different parts of the lensing galaxy[16,17]. Since Faraday effects intrinsic to the background source and that produced in the Milky Way foreground can be assumed to have negligible differences for all lensed images (see Methods), differential Faraday rotation and depolarization between lensed images A and B carry information on large- and small-scale magnetic fields, respectively, in the lensing galaxy. Unlike the standard Faraday rotation measure grid technique, the lensing approach can deliver Faraday rotation and Faraday dispersion produced by a distant intervening galaxy free of contamination from the background source and the Milky Way. Previous efforts to measure differential Faraday rotation of images in strongly lensed systems were not successful since they relied on narrowband radio polarization data that were difficult to interpret[17,18]. With broadband polarization data of CLASS B1152+199 and the rotation measure synthesis technique[19], we obtained Faraday depth spectra of the lensed images as shown in Fig. 2. Differences in the gas density and the line-of-sight projection of the large-scale magnetic field in the lensing galaxy along the two light paths produce a shift in the observer's frame of 510 rad m$^{-2}$ in the location of the main peak in the Faraday depth spectra. In addition, the presence of random magnetic fields and high gas column density along sightline B (ref. [14]) can lead to large Faraday rotation fluctuations on scales smaller than the telescope beam, which further depolarize and disperse incident background radiation. This can cause the broader peak and lower fractional polarization of the dominant Faraday component of image B compared with image A.

To quantify the appearance of the Faraday depth spectra and to extract physical properties of the magnetized gas in the lensing galaxy, we performed a direct fit to the observed fractional Stokes parameters $Q/I$ and $U/I$ as a function of frequency of the two images independently across 1-8 GHz. The complex fractional polarization spectrum $P$ of each lensed image can be best modelled by two distinct polarized components that have Faraday rotation and depolarization effects occurring external to synchrotron emitting regions[20] (see Methods)

$$P = \sum_{k=1}^{k=2} p_{0,k} e^{2i(\theta_{0,k} + \text{RM}_k c^2/\nu^2)} e^{-2\sigma_{\text{RM},k}^2 c^4/\nu^4} \quad (1) \, ,$$

where $\nu$ is the observing frequency, and $p_{0,k}$, $\theta_{0,k}$, $\text{RM}_k$, $\sigma_{\text{RM},k}$ are the intrinsic fractional polarization, intrinsic polarization angle, Faraday rotation and Faraday dispersion of the $k$-th components, respectively. The resulting best-fit parameters to the polarization spectra of images A and B are listed in Table 1 and the best-fit curves are displayed in Supplementary Fig 1. The two polarized components obtained by independent fits to the lensed images have intrinsic fractional polarization and polarization angles that differ by less than 1.4σ (see Methods). This strongly indicates that in polarization, we detected the same components of the background source in both lensed images. Therefore, differential magnification and time delay effects (this lensing system did not exhibit any variability through a six month radio monitoring campaign[21]), which alter the intrinsic polarization properties of the images, are unimportant in this lensing system and are not further considered in our analysis. The differential Faraday rotation (ΔRM) and differential Faraday dispersion ((Δσ$_{\text{RM}}^2$)$^{1/2}$ — the additional Faraday dispersion experienced by image B compared with image A) between the lensed images are clean *in situ* measures of the lensing galaxy's large-scale and small-scale magnetic field properties, respectively (see Methods). The lensing galaxy acts as a Faraday screen that produces rest-frame $|\Delta RM|= |\Delta RM_{obs}|(1+z)^2$ of $1040 \pm 60$ rad m$^{-2}$ and $(\Delta\sigma_{RM}^2)^{1/2}$ of $100 \pm 30$ rad m$^{-2}$ between the same polarized components of the two lensed images (computed using the more strongly polarized component), broadly similar in magnitude to the $|\Delta RM|$ and $(\Delta\sigma_{RM}^2)^{1/2}$ values reported in the Milky Way and in the nearby Large Magellanic Cloud[22-24]. The fact that the rest-frame $|\Delta RM|$ exceeds $(\Delta\sigma_{RM}^2)^{1/2}$ by a factor of ten is strong evidence for the presence of coherent magnetic fields: a purely random field characterized by an RM fluctuation of 100 rad m$^{-2}$ would only produce a small residual Faraday rotation after passing through a large number of turbulent cells along the line of sight (the expected RM after passing through N turbulent cells in a region without coherent fields is $\sim \sigma_{RM}/\sqrt{N}$ rad m$^{-2}$).

|  | $p_0$ (%) | $\theta_0$ (°) | RM (rad m$^{-2}$) | $\sigma_{RM}$ (rad m$^{-2}$) | $\chi_\nu^2$ |
|---|---|---|---|---|---|
| **Image A** |  |  |  |  | 1.18 |
| Component 1 | 2.89(7) | +32(1) | 0(1) | 13.2(8) |  |
| Component 2 | 1.1(1) | −38(4) | +160(20) | 61(9) |  |
| **Image B** |  |  |  |  | 1.02 |
| Component 1 | 1.9(7) | +30(10) | +500(30) | 50(10) |  |
| Component 2 | 1.0(8) | −20(20) | +710(80) | 70(40) |  |

*Table 1. Best-fit parameters and reduced $\chi^2$ of maximum likelihood fits to broadband polarization spectra of the lensed images in CLASS B1152+199. The best-fit model contains two distinct polarized components as described by equation (1). Uncertainties of the last quoted digit of the fitted parameters are listed in the parentheses. Note that intrinsic polarization angles $\theta_0$ reported in this table differ from those inferred from the main peak of the Faraday depth spectra due to an additional fitted polarized component.*

To estimate the large-scale magnetic field strength and geometry in the lensing galaxy, we constructed a model for the lensing galaxy's magnetized ISM. We assumed that the galaxy hosts a disk magnetic field which has either axisymmetric or bisymmetric geometry (anti-symmetric with respect to the rotational axis by 180°; ref. [25]), and the strength falls off exponentially with a radial scale length in the range of 5 kpc < $r_{MAG}$ < 20 kpc (ref. [1]). We assumed that the ionized gas column density towards image A through the lensing galaxy G ($N_{e,A} = \int_{\text{sightline A thru.G}} n_e \, dl$) ranges between 5 and 300 pc cm$^{-3}$ (this corresponds to the range of electron column density through a Milky Way-type galaxy with an inclination of 33°; ref. [26]). Furthermore, we adopted a differential electron column density ($N_{e,B}$-$N_{e,A}$) of 156 pc cm$^{-3}$ by assuming that 10% of the X-ray absorbing gas column is ionized (see Methods). The positions of the lensed images on the sky are converted into locations within the disk of the lensing galaxy using an inclination angle of 33° and a position angle of the line-of-node of −63°, derived from isophote fitting to the Hubble Space Telescope image[13] and an assumed intrinsic axial ratio for spiral galaxies of 0.15 (ref. [27]). We computed the range of plausible magnetic field strength in the lensing galaxy at the location of image B for values of $r_{MAG}$ and $N_{e,A}$ in the parameter space considered. Figure 3 shows the resulting magnetic field strength at the galacto-centric radius of image B (2.6 kpc) in the lensing galaxy for the axisymmetric and the bisymmetric field cases. The mean coherent magnetic field strength is 8 μG for the axisymmetric case and 11 μG for the bisymmetric case. We note that true field strengths could differ from the reported values by at most 40 % if we consider a range of possible ionization fractions within its 90% confidence range (see Methods). To account for the measured rest-frame differential RM for a specific $r_{MAG}$ and $N_{e,A}$, a bisymmetric field requires a higher field strength— up to a factor of 1.7 more— than an axisymmetric field.

While Faraday rotation is a measure of coherent magnetic fields, Faraday dispersion $\sigma_{RM}$ encodes random magnetic field information in the lensing galaxy on scales smaller than ~ 20 pc

— the projected lensed image size at the redshift of the lensing galaxy at milli-arcsec resolution[13]. Considering the case where only isotropic and homogenous fluctuations in magnetic fields contribute to $\sigma_{RM}$, the random magnetic field strength $B_r$ in the lensing galaxy on scales small than 20 pc is found to decrease with increasing $N_{e,A}$ from 6 μG to 3 μG, with a mean of 4 μG (see Methods).

According to arguments given in ref. [28], conservation of magnetic helicity imposes a lower limit of ~ 1.6 on the random-to-coherent magnetic field ratio ($B_r/B_{coherent}$) in galaxies for a dominant helicity scale of ~ 100 parsecs. If magnetic fluctuations in the lensing galaxy follow the Kolmogrov spectrum, the random field strength scales as $B_r \propto l^{1/3}$, and we can extrapolate $B_r$ from smaller than 20 pc scales and obtain $B_r$ on ~100 pc scales to be between 11 and 23 μG in the parameter space probed. For the axisymmetric field case, our observations imply a $B_r/B_{coherent}$ ratio that satisfies this criterion in over 40% of the parameter space. However, for the bisymmetric field case, our observations yield a ratio that falls below this limit across the entire parameter space considered. More generally speaking, since the estimated strength for a bisymmetric field exceeds that of an axisymmetric field, for any given random field strength, the helicity conservation criterion on $B_r/B_{coherent}$ can always be satisfied more easily by an axisymmetric field. Hence, an axisymmetric large-scale field in the lensing galaxy is favoured over a bisymmetric one.

We examined whether the mean-field dynamo is responsible for the observed coherent large-scale magnetic field in the lensing galaxy. Both the strength and the preferred field geometry in the lensing galaxy are compatible with a mean-field-dynamo-generated magnetic field. The mean-field-dynamo model predicts that the axisymmetric mode is the strongest mode excited in a galactic disk and that the random field is expected to dominate over the coherent field[29], as observed here. On-going star formation in this $z = 0.439$ galaxy, indicated by strong O[II] lines in its spectrum[12,30], and subsequent supernova explosions can drive turbulence and therefore the alpha-effect, which is a necessary ingredient for the mean-field dynamo[11,29]. Assuming Milky-Way-value dynamo parameters, the dynamo can produce fields of similar strength and coherency in a galaxy by $z = 0.439$ (ref. [31]).

We have presented direct measurements of μG magnetic fields with a significant coherent component in a galaxy when the universe was two-thirds of its current age. Magnetic field properties in this cosmologically distant galaxy have comparable strength and geometry to galaxies in the present day universe[1,8]. Since the mean-field dynamo has less time to operate in a young galaxy, the existence of coherent μG magnetic fields in this $z = 0.439$ galaxy has implications on the mean-field dynamo e-folding time $\tau_{dynamo}$: if the disk of the lensing galaxy settled into equilibrium at $z \sim 2$ (ref. [32]), and magnetic field saturation occurred before $z = 0.439$, then $\tau_{dynamo}$ has an upper limit of $\frac{5.8}{\ln(B_{z=0.439}/B_{seed})}$ Gyrs, where $B_{seed}$ is the seed magnetic field of the mean-field dynamo. For $B_{seed} \sim 3 \times 10^{-16}$ G (ref. [33]), $\tau_{dynamo} < 2.4 \times 10^8$ yrs, which is consistent with the typical rotational period of a spiral galaxy. Future such measurements of galactic magnetic fields at higher redshifts will provide a discriminant between the mean-field dynamo and other magnetic field generation mechanisms, such as the cosmic-ray driven dynamo[34].

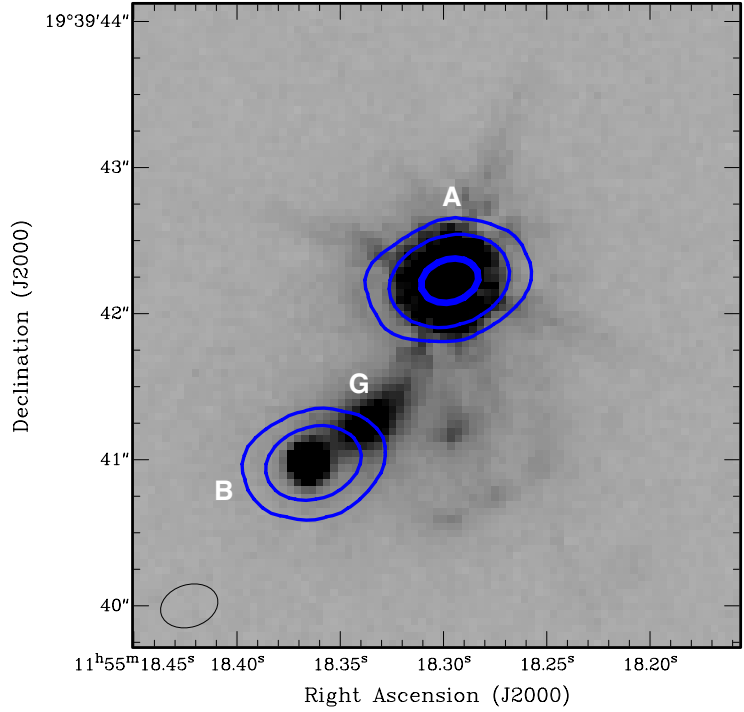

**Fig. 1. The 5 GHz total intensity radio contour of the gravitational lensing system CLASS B1152+199 overlaid on the HST F814W image.** The lensed images are labelled A and B, whereas the lensing galaxy is labelled G. The blue radio contours are at 0.2, 2.0 and 20.0 mJy beam$^{-1}$, respectively, with the lowest contour level corresponding to a 10σ detection. The resolution of the radio image is 0.4"×0.29", as indicated by the black ellipse in the lower left corner. Images A and B lie 1.1" and 0.45" away from the centre of the lensing galaxy G, which correspond to projected distances of 6.5 and 2.6 kpc for a cosmology with $H_0 = 69.6$ km s$^{-1}$ Mpc$^{-1}$, $\Omega_M = 0.286$, $\Omega_{vac} = 0.714$.

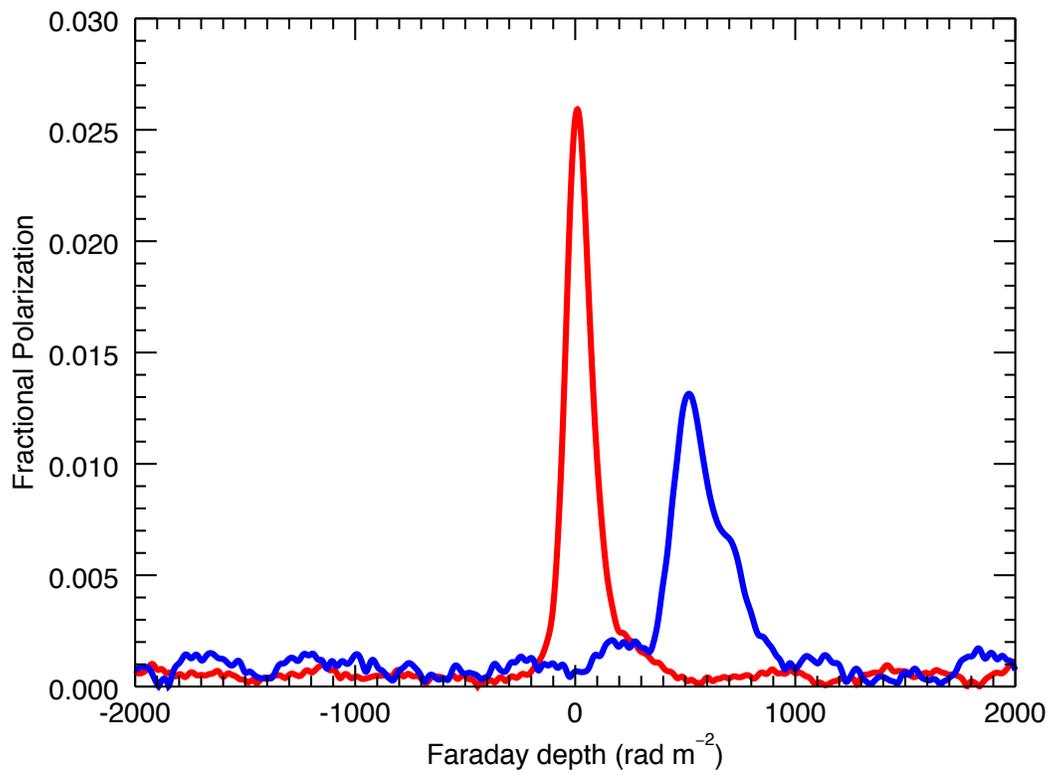

Fig. 2. Faraday depth spectra of images A (red) and B (blue) of the gravitational lensing system CLASS B1152+199 computed using the rotation measure synthesis technique,

**followed by deconvolution using the RM-Clean algorithm**[50]. The rms-noise levels in the spectra are $2\times10^{-4}$ and $7\times10^{-4}$ for images A and B, respectively. The dominant Faraday component of image A is located at $+9.7 \pm 0.5$ rad m$^{-2}$, and the dominant Faraday component of image B is located at $+517 \pm 3$ rad m$^{-2}$. The corresponding intrinsic polarization position angles (de-rotated to $\lambda = 0$ m) of image A ($+24.2° \pm 0.4°$) and image B ($+26° \pm 2°$) are consistent within their 1σ uncertainties, which strongly suggests that the dominant polarized components of the lensed images originate from the same part of the background source. The main Faraday component of image B is broader (full width at half maximum (FWHM) ≈ 191 rad m$^{-2}$) than that of image A (FWHM ≈ 129 rad m$^{-2}$) and it is about 50% less polarized than image A. This amount of additional broadening exceeds that inferred for circumgalactic medium of intervening galaxies[7], as our sightlines probe through the more turbulent galactic disk at small impact parameters. Finally, we note that a weaker Faraday component is visible in the positive tail of the main peak in both spectra.

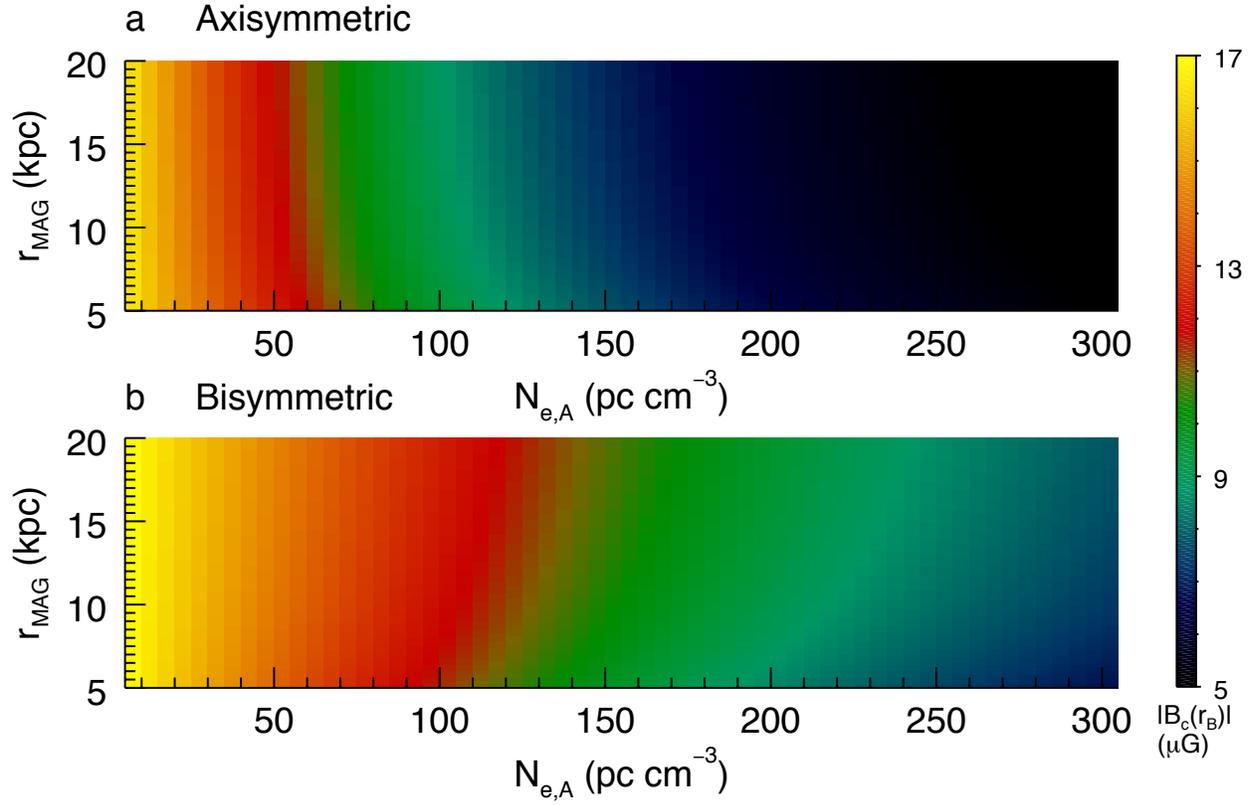

**Fig. 3. Coherent magnetic field strength ($|B_c(r_B)|$) at the galacto-centric radius of image B ($r_B$=2.6 kpc) in the lensing galaxy.** $|B_c(r_B)|$ is shown as a function of the electron column density along sightline A ($N_{e,A}$) and the radial magnetic field scale length $r_{MAG}$ for the case of an axisymmetric field (a) and a bisymmetric field (b) for 5 kpc < $r_{MAG}$ < 20 kpc and 5 pc cm$^{-3}$ < $N_{e,A}$ < 300 pc cm$^{-3}$. The magnetic field strength ranges between 4 and 16 μG for an axisymmetric field and between 7 and 17 μG for a bisymmetric field in the parameter space probed. The 1σ uncertainties associated with $|B_c(r_B)|$ range from 0.3 to 1 μG.

**Methods**

**Observations and Data Reduction.** The lensing system CLASS B1152+199 was observed on November 13th 2012 in the A array configuration with the VLA. Data were taken in L (0.994-2.006 GHz), S (1.988-4.012 GHz), and C bands (3.988-6.012 GHz, 5.988-8.012GHz), providing a continuous frequency coverage from 1 to 8 GHz. Data were recorded in 16 spectral windows in each band, with each spectral window (SPW) further divided into 64 1 MHz channels in L band and 64 2 MHz channels in S and C bands. The total on-source time across all bands was ~ 30 minutes. Data calibration and reduction were carried out using the Common Astronomy Software Applications (CASA)[35]. Visibilities affected by radio frequency interference were excised using the automated flagging algorithm 'rflag' in CASA with additional manual flagging. The standard VLA primary calibrator 3C286 was used for bandpass and absolute flux density calibrations. The absolute polarization angles were calibrated to 3C286 as well, whose angle is assumed to be constant (+33°) across 1-8 GHz (ref. [36]). Polarization leakage terms were determined from the unpolarized source J0713+4349. Using the secondary calibrator J1158+2450, we determined the time-dependent antennae gains of each SPW separately. Three rounds of phase-only self-calibration were performed on data in each SPW to reach the theoretical noise levels ~70 $\mu$Jy beam$^{-1}$, ~20 $\mu$Jy beam$^{-1}$ and ~15 $\mu$Jy beam$^{-1}$ across L, S and C bands respectively.

Imaging and deconvolution were performed using the CASA task 'clean'. Supplementary Fig. 2 shows the total intensity and polarized intensity images of the system centred at 5 GHz across a 2 GHz bandwidth. For the purpose of rotation measure synthesis and polarization analysis, we imaged Stokes $I$, $Q$, and $U$ of the lensing system averaged in 8 MHz channels between 1 and 8 GHz using robust weighting. The native resolutions of these channel maps range from 2.36"×1.75" at 1.006 GHz to 0.23"×0.22" at 7.991 GHz. Since both lensed images remain unresolved at the highest observing frequency, the two lensed images were subsequently fitted as point sources in each 8 MHz channel map in all Stokes using the MIRIAD[37] task 'imfit'. We note that since images A and B are separated by 1.56", they are blended at the lowest frequencies. If one followed the standard practice of smoothing all channel maps down to the angular resolution of the lowest frequency channel map before polarization analysis, it would lead to the loss of spatial information of the corresponding Faraday components. We chose to fit for the flux densities at the native resolution of the channel maps to retain spatial information of the Faraday components.

**Faraday Rotation Synthesis.** The frequency coverage of 1-8 GHz with 8-MHz-wide channels yields a FWHM of the rotation measure spread function of 110 rad m$^{-2}$. The channel width limits our sensitivity to Faraday depths with a magnitude of less than $6\times10^5$ rad m$^{-2}$. The highest observing frequency sets our sensitivity to extended structures in Faraday depth space: the sensitivity drops by 50% for structures with extents greater than 2230 rad m$^{-2}$ (ref. [19]).

**Stokes *QU* fitting.** Since extracting properties of the underlying Faraday structures based on RM synthesis alone has considerable ambiguities[38,39], we directly fitted the observed fractional Stokes *Q* and *U* as a function of frequency to various models of the line-of-sight synchrotron-emitting and Faraday-rotating magnetized gas to reliably extract properties of the magneto-ionic medium being probed[40].

The lensed polarized source itself is modeled as being either Faraday thin (polarized emission emitted at a single Faraday depth) or Faraday thick (polarized emission emitted at multiple Faraday depths from either a combination of multiple Faraday thin components or polarized emissions that have been Faraday dispersed and depolarized). Specifically, we considered the following models for the background source: (1) a single Faraday thin component; (2) a single polarized component propagating through an inhomogeneous external Faraday screen; (3) two spatially unresolved Faraday thin components; (4) two spatially unresolved polarized components propagating through two different inhomogeneous external Faraday screens; (5) a Burn slab, which consists of regular fields and well-mixed thermal and cosmic-ray electrons producing Faraday depolarization; (6) a slab with both regular and random magnetic fields in a volume of well-mixed thermal and cosmic-ray electrons; and (7) three spatially unresolved Faraday thin components.

The complex polarized emission from the lensed source subsequently propagates through the lensing galaxy's ISM, acting as an external Faraday screen that could produce additional Faraday rotation and depolarization because of the presence of coherent and small-scale magnetic fields. Finally, this polarized emission penetrates the magneto-ionic medium of the Milky Way, producing further Faraday rotation.

We assumed that the intergalactic medium contributes a negligible amount to the observed Faraday rotations[41] and that the variation of the Milky Way foreground Faraday rotation on arcsec scale is negligible[42]. In addition, we assumed that random magnetic fields in the inhomogeneous Faraday screens fluctuate on scales much smaller than the telescope beam, such that depolarization effects can be described analytically. We note that isolating RM contributions from Faraday screens associated with different parts of a sightline is not possible even with Stokes $QU$ fitting because one effectively measures a single net Faraday rotation and external Faraday dispersion from all the screens. This is the precise reason why we require closely spaced sightlines provided by strong gravitational lensing systems to separate out the Faraday rotation and Faraday dispersion produced in the lensing galaxy, as Faraday effects produced by the background source and the Milky Way have negligible differences for all lensed images.

We used Faraday components identified in the deconvolved Faraday depth spectrum as initial guesses for the fitted parameters. We decided on the best-fit model by comparing the reduced $\chi^2$ of the models using the f-test. We selected the least complex model that minimizes the reduced $\chi^2$: a more complex model is only favoured when the f-test suggests that it improves the fit by at least 2.3σ (~ 98% confidence level) over the simpler model. A lensed polarized source described by model (4) including effects of the lensing galaxy and the Milky Way with complex polarization in the form of equation (1) was found to provide the best fit to the observed broadband polarization data. The Stokes $QU$ data and the best-fit curves to the broadband polarized spectra of images A and B are presented in Supplementary Fig. 1.

We note that unlike narrowband polarization data, which can suffer from nπ ambiguity (where multiple rotation measure values can provide equal good fits to the data), our broadband polarization data with fine frequency sampling can eliminate this problem, thus providing unique fits to the observations.

Finally, we repeated Stokes $QU$ fitting using only data between 2 and 8 GHz, as blending of the lensed images below 2 GHz may affect the results. We found the best-fit model and the corresponding best-fit parameters to be consistent within uncertainties with results obtained using 1-8 GHz data. Hence, blending of the lensed images at low frequencies does not affect our Stokes $QU$ fitting results.

**Polarization properties of the lensed background source.** Independent Stokes $QU$ fits to the lensed images A and B yield consistent polarization properties for the lensed background source: the best-fit values for $p_{0,\,\text{component 1}}$ are different at the 1.4σ level, whereas $\theta_{0,\,\text{component 1}}$, $p_{0,\,\text{component 2}}$ and $\theta_{0,\,\text{component 2}}$ are all consistent within their uncertainties. In addition, the derived Faraday rotation difference of the two polarized components ($\text{RM}_{\text{component 1}} - \text{RM}_{\text{component 2}}$) for image A is in excellent agreement with that obtained for image B, which further strengthens the argument that we have detected the same components of the background source in both lensed images in polarization.

**Probe of the large-scale magnetic fields.** As both polarized components inferred from Stokes *QU* fitting experience the same amount of differential Faraday rotation and dispersion (which is not expected if the light bundle passes through clumpy turbulent local features such as an HII region), the observed differential Faraday rotation and dispersion is most likely produced by coherent magneto-ionic structures on large scales. Even in the unlikely case that the light path through image B does penetrate an HII region, the magnetic field strength and direction inferred from Faraday rotation would still be representative of the galactic-scale fields in the lensing galaxy, as have been shown for Galactic HII regions[43].

**Estimating the differential electron column density.** To convert the differential hydrogen column density[14] $|\Delta N_H| = (0.48 \pm 0.04) \times 10^{22}$ cm$^{-2}$ into a differential electron column density, we adopted an ionization fraction of 10%, the Galactic empirical value determined by comparing X-ray absorption column densities and dispersion measures towards radio pulsars[44]. This resulted in a differential electron column density $\Delta N_e = N_{e,B} - N_{e,A}$ of $156 \pm 13$ pc cm$^{-3}$. We note that this value implicitly takes into account the free electrons provided by helium ionization. The ionization fraction used for the conversion is reasonable for our lensing system, which qualifies as a damped Lyman-α system[15] (a quasar absorption line system with HI column density exceeding

$2\times10^{20}$ cm$^{-2}$). Since the ionization fraction in sub-damped Lyman-α systems at $z \sim 0.5$ is found to be less than 15% (ref. [45]), damped Lyman-α systems in the same redshift range are expected to be even less ionized due to self-shielding[46]. Based on Cosmic Origin Spectrograph observations, the system has recently been identified as a ghostly damped Lyman-α system, where the expected Lyman-α absorption trough is filled by Lyman-α emission from the lensing galaxy[47]. The uncertainty in the adopted ionization fraction $10^{+4}_{-3}\%$ (ref. [44], 90% confidence level) has the following impact on the estimated magnetic field strengths: if the ionization fraction is 7%, the reported field strengths in this paper will be a factor of 1.4 higher, whereas if the ionization fraction is 14%, the reported field strengths in this paper will be a factor of 0.7 lower.

**Determining the large-scale coherent magnetic field strength.** We constructed a model for the magneto-ionic medium in the lensing galaxy in order to translate the rest-frame differential Faraday rotation into a magnetic field strength estimate. We assumed that the lensing galaxy hosts large-scale magnetic fields of either axisymmetric or bisymmetric geometry, with its strength $B_c(r)$ following an exponential with radial scale length $r_{\mathrm{MAG}}$: $B_c(r) = B_0 \exp(-r/r_{\mathrm{MAG}})$. Given a value of the electron column density along sightline A $N_{e,A}$ between 5 and 300 pc cm$^{-3}$, the electron column density along sightline B was determined by $N_{e,B}=\Delta N_e+N_{e,A}$. In the thin disk approximation, and under the assumption that magnetic fields and electron densities are uncorrelated, the differential Faraday rotation between images A and B can be expressed as $\Delta\mathrm{RM} \approx k[N_{e,B}B_{\parallel,B} - N_{e,A}B_{\parallel,A}]$, where $k = 0.812$ rad m$^{-2}$ pc$^{-1}$ cm$^3$ μG$^{-1}$ and $B_{\parallel,A}$ and $B_{\parallel,B}$ are line-of-sight projection of the lensing galaxy's magnetic field in μG along images A and B respectively. The coherent disk magnetic field strength $B_0$ can be estimated by

$$B_0 \approx \frac{\Delta \mathrm{RM}}{k \sin i \, | C_1 \mathrm{N}_{e,A} e^{-\frac{r_A}{r_{\mathrm{MAG}}}} - C_2 \mathrm{N}_{e,B} e^{-\frac{r_B}{r_{\mathrm{MAG}}}} |}.$$

Here, $i$ is the inclination of the disk of the lensing galaxy, while $C_1$ and $C_2$ are constants whose values depend on the field geometry.

For an axisymmetric field, $C_1 = \cos p_0 \cos\theta_A + \sin p_0 \sin\theta_A$ and $C_2 = \cos p_0 \cos\theta_B + \sin p_0 \sin\theta_B$, where $p_0$ is the pitch angle adopted to be −20°, typical for galaxies observed in the local volume[1,48,49] and $\theta_A$ and $\theta_B$ are the azimuthal angles of the sightlines A and B in the frame of the lensing galaxy. For a bisymmetric field,

$$C_1 = \cos p_1 \cos\theta_A \cos(\theta_A - \beta_1) + \sin p_1 \sin\theta_A \cos(\theta_A - \beta_1)$$

and

$C_2 = \cos p_1 \cos\theta_B \cos(\theta_B - \beta_1) + \sin p_1 \sin\theta_B \cos(\theta_B - \beta_1)$, where the $p_1$ is adopted to be −20°, and $\beta_1$, which determines the azimuth where the bisymmetric mode is maximum, is assumed to be 0° so that it gives the lower limit of the bisymmetric field strength $B_0$.

**Determining the small-scale random magnetic field strength.** For isotropic and homogenous magnetic field fluctuations, the random magnetic field strength $B_r$ in the lensing galaxy on scales smaller than 20 pc can be estimated using[1]

$$B_r = \frac{\sqrt{3fN\Delta\sigma_{\mathrm{RM}}^2}}{k\sqrt{\mathrm{N}_{e,B}^2 - \mathrm{N}_{e,A}^2}} = \frac{1.02 \times 10^3}{\sqrt{\mathrm{N}_{e,B}^2 - \mathrm{N}_{e,A}^2}} \sqrt{\frac{(f/0.05)(L/1 \text{ kpc})}{(l/2\text{pc})}} \left( \frac{\sqrt{\Delta\sigma_{\mathrm{RM}}^2}}{100 \text{ rad m}^{-2}} \right) \mu G.$$

We assumed a filling factor $f \approx 0.05$ and a path length $L$ of 1 kpc through the lensing galaxy such that for a turbulent cell size $l \sim 2$ pc a sightline contains $N \sim 5 \times 10^2$ cells. This yields a random magnetic field strength ranging between 3 to 6 μG for 5 pc cm$^{-3}$ < $N_{e,A}$ < 300 pc cm$^{-3}$. The dependence of $B_r$ on the assumed path length $L$ is not strong: a factor of four difference in the assumed $L$ will only result in a factor of two difference in $B_r$.

# Supplementary Figures

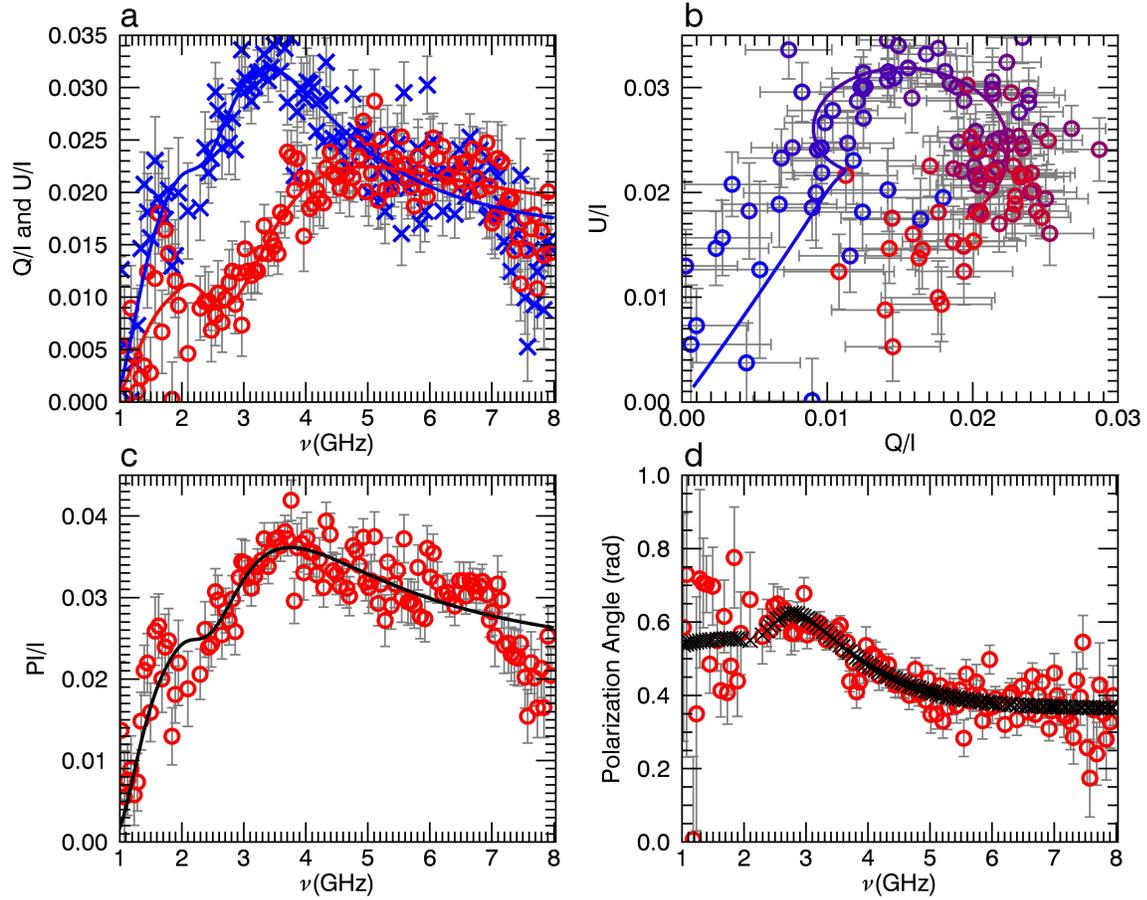

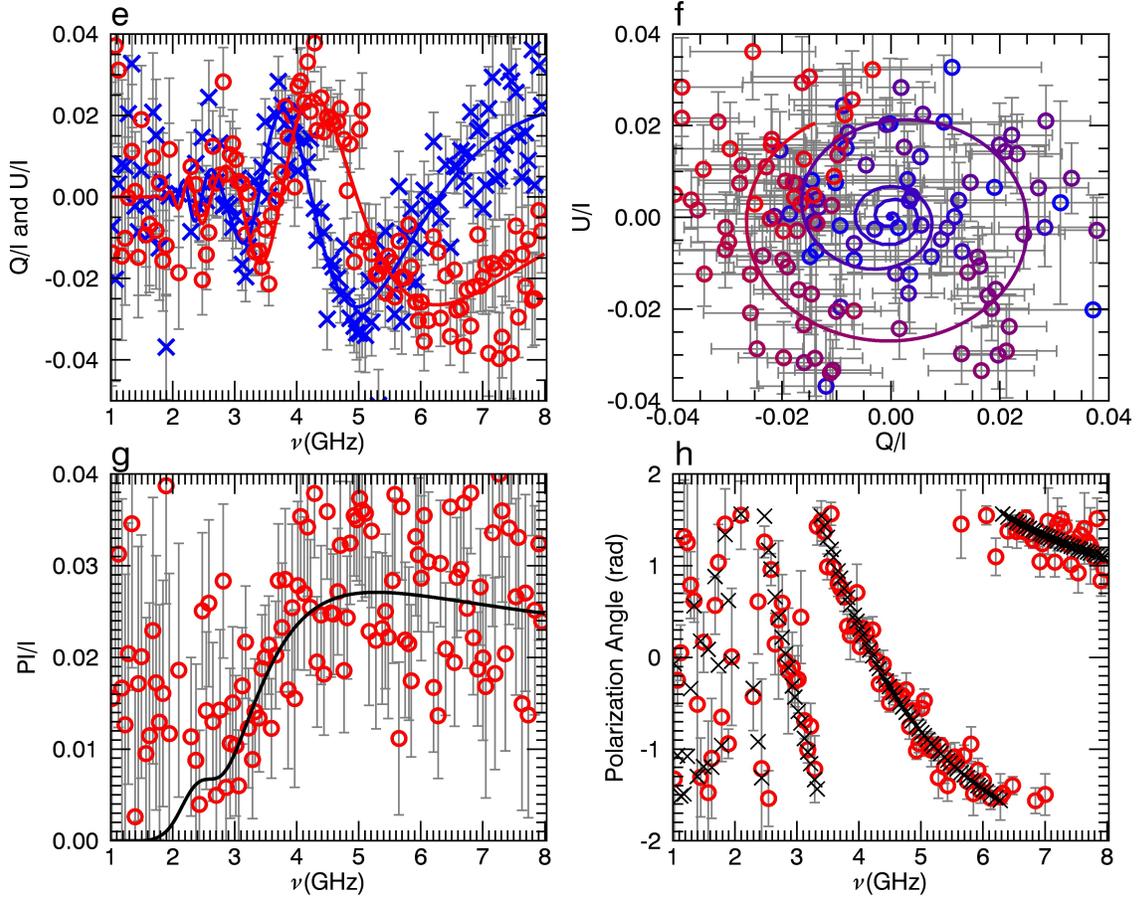

**Supplementary Figure 1 — Polarized emission of image A (panels a-d) and image B (panels e-h) across 1-8 GHz and the corresponding best-fits to model (iv).** For each lensed image, the observed fractional Q (red circles), fractional U (blue crosses) and their corresponding best-fit curves (solid lines) against frequency are displayed in panels a and e. In addition, the data (open circles) and the best fit model (solid line) in the U/I-Q/I track are plotted in panels b and f, both are color-coded such that the color gradient varies from blue to red with increasing frequency from 1-8 GHz. Fractional polarization P/I against ν is plotted in panels c and g, where red open circles are the data points and the black solid line represents the best-fit. Finally, the variation of the measured polarization position angles in radians across 1-8 GHz is plotted in panels d and h: red open circles are the data points and black crosses represent the best-fit values. For clarity, data displayed here have been averaged over 40 MHz. In all panels, 1σ error bars of the binned data are plotted in gray.

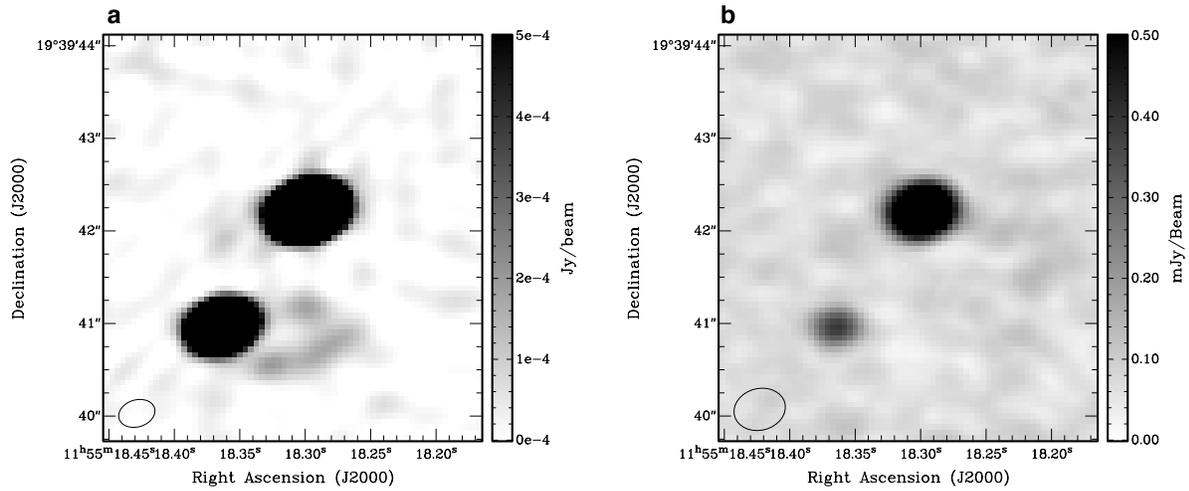

**Supplementary Figure 2 — The band-averaged total intensity (panel a) and polarized intensity (panel b) image of CLASSB1152+199 across 4-6 GHz.** The beam size for the total intensity and the polarized intensity images, as indicated by the black ellipse in the lower left corner of each image, are 0.4"×0.29" and 0.56" ×0.45", respectively.

**Acknowledgments:** The National Radio Astronomy Observatory is a facility of the National Science Foundation operated under cooperative agreement by Associated Universities, Inc.

Correspondence and requests for materials should be addressed to S. A. M.

**References**

1. Beck, R. Magnetic fields in spiral galaxies. *Annu. Rev. Astron. Astrophys.* **24**, 4-61 (2016).

2. Kronberg, P. P., Perry, J. J. & Zukowski, E. L. H. Discovery of extended Faraday rotation compatible with spiral structure in an intervening galaxy at *z* = 0.395 - new observations of PKS 1229 – 021. *Astrophys. J.* **386**, 528-535 (1992).

3. Oren, A. L. & Wolfe, A. M. A Faraday rotation search for magnetic fields in quasar damped Ly alpha absorption systems. *Astrophys. J.* **445**, 624-641 (1995).

4. Joshi, R. & Chand, H. Dependence of residual rotation measure on intervening Mg II absorbers at cosmic distances. *Mon. Not. R. Astron. Soc.* **434**, 3566-3571 (2013).

5. Bernet, M. L., Miniati, F., Lilly, S. J., Kronberg, P. P. & Dessauges-Zavadsky, M. Strong magnetic fields in normal galaxies at high redshift. *Nature* **454**, 302-304 (2008).

6. Farnes, J. S., O'Sullivan, S. P., Corrigan, M. E. & Gaensler, B. M. Faraday rotation from magnesium II absorbers toward polarized background radio sources. *Astrophys. J.* **795**, 63-89 (2014).

7. Kim, K. S. *et al*. Faraday Rotation Measure Synthesis of Intermediate Redshift Quasars as a Probe of Intervening Matter. *Astrophys. J.* **829**, 133-155 (2016).

8. Beck, R. & Wielebinski, R. in *Planets, Stars and Stellar Systems Vol. 5: Galactic Structure and Stellar populations* (eds Oswalt, T. D. & Gilmore, G.) 641-723 (Springer, Dordrecht, the Netherlands, 2013).

9. Kronberg, P. P. *Cosmic Magnetic Fields* (Cambridge University Press, 2016).

10. Ruzmaikin, A. A., Shukurov, A. M. & Sokoloff, D. D. *Magnetic Fields of Galaxies* (Kluwer, Dordrecht, the Netherlands, 1988).

11. Kulsrud, R. M. & Zweibel, E. G. On the origin of cosmic magnetic fields. *Rep. Prog. Phys.* **71**, 046901 (2008).

12. Myers, S. T. *et al.* CLASS B1152+199 and B1359+154: Two new lens systems discovered in the cosmic lens all-sky survey. *Astrophys. J.* **117**, 2565-2572 (1999).

13. Rusin, D. *et al.* High-resolution observations and mass modeling of the CLASS gravitational lens B1152+199. *Mon. Not. R. Astron. Soc.* **300**, 205-211 (2002).

14. Dai, X. & Kochanek, C. S. Differential X-ray absorption and dust-to-gas ratios of the lens galaxies SBS 0909+523, FBQS 0951+2635, and B1152+199. *Astrophys. J.* **692**, 677-683 (2009).

15. Toft, S., Hjorth, J. & Burud, I. The extinction curve of the lensing galaxy of B1152+199 at *z* = 0.44. *Astron. Astrophys.* **357**, 115-119 (2000).

16. Dyer, C. C. & Shaver, E. G. On the rotation of polarization by a gravitational lens. *Astrophys. J. Letters.* **390**, L5-L7 (1992).

17. Narasimha, D. & Chitre, S. M. Large scale magnetic fields in lens galaxies. *J. Korean Astron. Soc.* **37**, 355-359 (2004).

18. Patnaik, A. R., Menten, K. M., Porcas, R. W. & Kemball. A. J. in *Gravitational Lensing: Recent Progress and Future Goals* (eds Brainerd T. G. & Kochanek C. S.) 99-100 (Astronomical Society of the Pacific, San Francisco, 2001).

19. Brentjens, M. A. & de Bruyn, A. G. Faraday rotation synthesis. *Astron. Astrophys.* **441**, 1217-1228 (2005).

20. Sokoloff, D. D. *et al*. Depolarization and Faraday effects in galaxies. *Mon. Not. R. Astron. Soc.* **299**, 189-206 (1998).

21. Rumbaugh, N. *et al*. Radio monitoring campaigns of six strongly lensed quasars. *Mon. Not. R. Astron. Soc.* **450**, 1042-1056 (2015).

22. Brown, J. C. *et al*. Rotation measures of extragalactic sources behind the southern Galactic plane: new insights into the large-scale magnetic field of the inner Milky Way. *Astrophys. J.* **663**, 258-266 (2007).

23. Haverkorn, M., Brown, J. C., Gaensler, B. M. & McClure-Griffiths, N. M. The outer scale of turbulence in the magnetoionized Galactic interstellar medium. *Astrophys. J.* **680**, 362-370 (2008).

24. Gaensler, B. M. *et al*. The magnetic field of the Large Magellanic Cloud revealed through Faraday rotation. *Science* **307**, 1610-1612 (2005).

25. Beck, R., Brandenburg, A., Moss, D., Shukurov, A. & Sokoloff, D. Galactic Magnetism: Recent Developments and Perspectives. *Annu. Rev. Astron. Astrophys.* **34**, 155-206 (1996).

26. Xu, J. & Han, J. L. Extragalactic dispersion measures of fast radio bursts. *Research Astro. & Astrophys.* **15**, 1629-1638 (2015).

27. Masters, K. L. *et al*. Galaxy Zoo: dust in spiral galaxies. *Mon. Not. R. Astron. Soc.* **404**, 792- 810 (2010).

28. Shukurov, A. in *Mathematical Aspects of Natural Dynamos* (eds Dormy, E. A. & Soward, M.) Ch. 7 (CRC Press, Boca Baton, FL, 2007).

29. Shukurov, A. in *Cosmic Magnetic Fields* (eds Wielebinski, R. & Beck, R) 113-135 (Springer Berlin Heidelberg, Berlin, Germany 2005).

30. Momcheva, I. G., Williams, K. A., Cool, R. J., Keeton, C. R. & Zabludoff, A. I. A spectroscopic survey of the fields of 28 strong gravitational lenses: the redshift catalog. *Astrophys. J. Suppl.* **219**, 29-61 (2015).

31. Arshakian, T. G., Beck, R., Krause, M. & Sokoloff, D. Evolution of magnetic fields in galaxies and future observational tests with the Square Kilometre Array. *Astron. Astrophys.* **494**, 21-32 (2009).

32. van der Kruit, P. C. & Freeman, K. C. Galaxy Disks. *Annu. Rev. Astron. Astrophys.* **49**, 301-371 (2011).

33. Neronov, A. & Vovk, I. Evidence for strong extragalactic magnetic fields from Fermi Observations of TeV Blazars. *Science* **328**, 73-75 (2010).

34. Hanasz, M., Wóltański, D. & Kowalik, K. Global galactic dynamo driven by cosmic rays and exploding magnetized stars, *Astrophys. J.* **706**, L155-L159 (2009).

35. McMullin, J. P., Waters, B., Schiebel, D., Young, W. & Golap, K. in *Astronomical Data Analysis Software and Systems XVI* (eds Shaw, R. A., Hill, F. & Bell D. J.) 127-130 (Astronomical Society of the Pacific, San Francisco, CA, 2007).

36. Perley, R. A. & Butler, B. J. Integrated polarization properties of 3C48, 3C138, 3C147, and 3C286. *Astrophys. J. Suppl.* **206**, 16-23 (2013).

37. Sault, R. J., Teuben, P. J & Wright, M. C. H. "A retrospective view of MIRIAD" in *Astronomical Data Analysis Software and Systems IV*. Shaw, R. A., Payne, H. E. & Hayes, J. J. E. Eds. (ASP, 1995), vol. **77**, pp 433-436.

38. Farnsworth, D., Rudnick, L. & Brown, S. Integrated polarization of sources at $\lambda \sim 1$ m and new rotation measure ambiguities. *Astron. J.* **141**, 191-219 (2011).

39. Sun, X. H. *et al*. Comparison of algorithms for determination of rotation measure and

Faraday structure. I. 1100-1400 MHz. *Astron. J.* **149**, 60-73 (2015).

40. O'Sullivan, S. P. *et al*. Complex Faraday depth structure of active galactic nuclei as revealed by broad-band radio polarimetry. *Mon. Not. R. Astron. Soc.* **412**, 3300-3315 (2012).

41. Akahori, T. & Ryu, D. Faraday rotation measure due to the Intergalactic Magnetic Field. II. The Cosmological Contribution. *Astrophys. J.* **738**, 134-142 (2011).

42. Leahy, J. P. Small-scale variations in the Galactic Faraday rotation. *Mon. Not. R. Astron. Soc.* **226**, 433-446 (1987).

43. Harvey-Smith, L., Madsen, G. J. & Gaensler, B. M. Magnetic Fields in Large-diameter HII Regions Revealed by the Faraday Rotation of Compact Extragalactic Radio Sources. *Astrophys. J.* **736**, 83-95 (2011).

44. He, C., Ng, C.-Y. & Kaspi, V. M. The correlation between dispersion measure and X-ray column density from radio pulsars. *Astrophys. J.* **768**, 64-72 (2013).

45. Peroux, C., Dessauges-Zavadsky, M., D'Odorico, S., Kim, T. S. & McMahon, R. G. A homogenous sample of sub-damped Lyman α systems–IV. Global metallicity evolution. *Mon. Not. R. Astron. Soc.* **382***,* 117-193 *(*2007).

46. Wolfe, A. M., Gawiser, E. & Prochaska, J. X. Damped Lyα systems. *Annu. Rev. Astron. Astrophys.* **43**, 861-918 (2005).

47. Dai, X. & Chen, B. Identifying the Lens Galaxy B1152+199 as a Ghostly Damped Lyman Alpha System by the Cosmic Origin Spectrograph. Preprint at **https://arxiv.org/abs/1612.04848** (2016).

48. Fletcher, A. in *The Dynamic Interstellar Medium: a Celebration of the Canadian Galactic Plane Survey* (eds Kothes, R., Landecker, T. L. & Willis, A. G.) 197-210 (Astronomical Society of the Pacific, San Francisco, CA, 2010).

49. Van Eck, C. L., Brown, J. C., Shukurov, A. & Fletcher, A. Magnetic Fields in a Sample of Nearby Spiral Galaxies. *Astrophys. J.* **799**, 35-54 (2015).

50. Heald, G., Braun, R. & Edmonds, R. The Westerbork SINGS survey. II Polarization, Faraday rotation, and magnetic fields. *Astron. Astrophys.* **503**, 409-435 (2009).